\documentclass{article}
\usepackage{graphicx,graphics}
\usepackage{booktabs}
\usepackage{epsfig}

\usepackage{amssymb}
\usepackage{amsthm}

\usepackage{float}
\usepackage{amssymb,amsmath,isomath,amsfonts,subfig}

\begin{document}

\title{Structural Changes of Active Skeletal Muscles: Modelling, Validation and Numerical Experiments}

\author{Hadi Rahemi$^{*,1}$, Nilima Nigam$^2$, James M. Wakeling$^1$\\
{\small 1. Dept. of Biomechanics and Kinesiology, Simon Fraser University}\\
{\small 2. Dept. of Mathematics, Simon Fraser University}}

\maketitle

%\address[bpk]{Neuromuscular Mechanics Laboratory, Department of Biomedical Physiology and Kinesiology, Simon Fraser University, Burnaby, BC, Canada V5A 1S6}
%\address[math]{Department of Mathematics, Simon Fraser University, Burnaby, BC, Canada}

\begin{abstract}
 The purpose of this study was to test some validation simulations for a 3D finite element model of contracting muscle. The model was based on continuum theory for fibre-reinforced composite materials. Here we simulated contractions for an idealized medial gastrocnemius muscle in man, using the model. Simulations were performed to test the force-length relation of the whole muscle, to evaluate the changes in internal fascicle geometry during contractions, and to assess the importance of material formulations for the aponeurosis and tendon. The simulation results were compared to previously published experimental values. The force-length profile for the whole muscle showed a realistic profile. As the muscle contracted the fascicles curved into S-shaped trajectories and curled around 3D paths, both of which matched previous experimental findings. As the fascicles shortened they increased in their cross-sectional area, but this increase was asymmetric with the smaller increase occurring within the fascicle-plane: the Poisson's ratio in this plane matched that previously shown from ultrasound imaging. The distribution of strains in the aponeurosis and tendon was shown to be a function of their material properties. This study demonstrated that the model could replicate realistic patterns of whole muscle-force, and changes to the internal muscle geometry, and so will be useful for testing mechanisms that affect the structural changes within contracting muscle. 
\end{abstract}

{\bf Keywords:} Skeletal muscle model, finite element method, model validation, fascicle pennation, fascicle curvature, connective tissue properties 

\section{Introduction}
Forces developed by contracting skeletal muscle depend on the structure and geometry of the contracting fascicles, and their interaction with the surrounding connective tissues. Recent studies have highlighted the complexity of the internal structure of the muscles in 3D, and the changes to this structure during contraction e.g. \cite{Rana20133d}. However, relatively little is known about the mechanisms that relate the structure to function. It is likely that regional variations in muscle structure, tissue properties and activation patterns all contribute to the force output from the muscle. In order to understand such effects it is necessary to use a muscle model that can incorporate these complexities. An efficient way, in terms of both time and cost, to test these effects would be with a 3D finite element simulation platform based on a realistic mathematical model of muscle.

Muscle models and their related simulations have evolved over the last decade to incorporate 3D structural and architectural parameters such as fascicle orientations and connective tissue properties e.g.,\cite{Oomens2003,Blemker2005,Bol2008}. Features such as fascicle activation patterns, structural changes (for instance changes in fascicle curvature and orientation) under isometric and dynamic contractions and their effects on the force and power generated by the whole muscle have been investigated in a number of previous works e.g.\cite{rahemi2014regionalizing, Carrasco1999}. While recent developments in imaging and signal processing techniques are enhancing our ability to measure detailed structure \cite{namburete2011computational, Rana20133d} and activation profiles e.g.\cite{hodson2013myoelectric,kinugasa2011unique,staudenmann2009heterogeneity} in a muscle, all the intended parameters may be hard or impossible to collect in a single experiment. Therefore, there is a need to use mathematical models to get insight into muscle function where large number of parameters can be manipulated or measured during a simulation of muscle contraction.

Here we present the results of 3D finite element simulations of a skeletal muscle model that has been developed specifically to investigate the relation between the muscle’s internal structure and activation patterns and its force output \cite{rahemi2014regionalizing}. The model has the ability to include detailed 3D architecture and regionalized submaximal activity in different groups of fascicles. It integrates the effects of different tendon and aponeurosis properties on the force transfer within the muscle-tendon unit from its origin to insertion. Furthermore, we have previously shown that this mathematical modelling framework can predict the deformations of the internal structure within the muscle, and the force vector developed by the whole muscle, while the activity patterns within the muscle can be varied and regionalized \cite{rahemi2014regionalizing}.

The main purpose of the current work is to present the validity of this modelling framework using different sets of experimental data. A validated computational model of muscle can be used to test mechanisms and investigate the effect of parameters that are difficult or impossible to measure. The second purpose of this work is to demonstrate some of the effects of the tendon and aponeurosis properties on the structural properties of the muscle during contraction. 

\section{Methods}

A 3D finite-element model of MG muscle was developed based on the continuum theory for fibre-reinforced composite materials. The tissues were transversely isotropic and were constrained to have nearly incompressible behaviour. The mathematical framework for this work has been previously described \cite{rahemi2014regionalizing}. The computational model was validated by comparing the force-length properties of the whole muscle to experimental measures, and also by comparing the shape, orientation and curvature of the modelled muscle fascicles to similar measures that have recently been made available through ultrasound imaging studies. 

A unipennate muscle belly was modelled with dimensions similar to the medial gastrocnemius in man. The model coordinate system had the Z-axis running proximal-distal along the line of action of the muscle, the Y-axis ran from the deep to the superficial direction and the X-axis ran across the medial-lateral width of the muscle. This model had the same constitutive law and geometries that  we have previously used \cite{rahemi2014regionalizing}. However, for this study the activation patterns and structural parameters along with mathematical boundary and initial condition were altered. The end planes of aponeuroses were defined as the transverse planes where the aponeuroses would join onto the external tendons, and mark the proximal and distal ends of the muscle belly. Some simulations were run for isometric contractions of the muscle belly where the end planes of the aponeuroses were fixed. Other simulations were run for the whole muscle-tendon unit with the external tendons included: for these, the proximal and distal ends of the muscle-tendon unit were fixed during contraction. 

Simulations in this study were done using a set of C++ libraries for finite element modelling \cite{dealii}. Each simulation was run with an increasing and uniform level of activation across all fascicles. The simulations were terminated when the nonlinear iterations did not converge within specified tolerences within given number of steps; this point depended on the initial state and boundary conditions for each simulation. Where groups of simulations are compared together, they were compared up to the highest activation level that was commonly achieved across the set. Each simulation took approximately 10 minutes to run [on a standalone 8-core (16 thread) computer], and this time included that for mesh initialization, matrix setup, iterative solving and result output.

\subsection{Simulation vs. Experiments - Validation of a muscle model}
Two sets of simulations were carried out on a muscle belly geometry (see Figure 1 in \cite{rahemi2014regionalizing}). Initially the muscle belly was a parallelepiped
 with 65 mm initial fascicle length, 15 degree pennation angle, and each aponeurosis was a rectangular cuboid of 210 $\times$ 55 $\times$ 3 mm. The initial stretch values for both the muscle and aponeuroses fascicles were set to one.
This stretch corresponds to the optimal length for the muscle fascicles. A set of simulations was run to map the force-length relation for the muscle belly, and a second set of simulations was run to test the trajectories of the muscle fascicles and the strains within the tissues during contraction.

\subsubsection{Force-length test for isometric contractions of a muscle belly}
The model of the muscle belly was adjusted to different lengths by fixing one end at its aponeurosis end plane, and passively displacing the other aponeusosis end plane to a new position. When the length of the muscle belly reached the desired length, both end planes for the aponeuroses were fixed to maintain the muscle belly at an isometric length, and the activation level in the muscle fascicles was then ramped up. The
range over which the muscle belly length changed was selected so that pre-activation fascicle stretch in the fascicles was between 0.75 and 1.35. This is close to the range for stretches in the medial gastrocnemius (MG) that have been reported when the 
ankle is passively moved from 30 degrees plantarflexion to 15 degrees dorsiflexion \cite{Maganaris1998}. To achieve this, the muscle belly
was shortened about 6\% for the lower bound of the fascicle stretch range. However, lengthening of the belly was selected to surpass the natural range so the force-stretch curve could be plotted for a longer range. The simulations at different lengths reached a common activity level of 30\%. The magnitude of the passive and total belly forces were computed along with the muscle fascicle lengths at which those forces were developed. The active muscle force was taken as the difference between the total force and the passive force for a set of common muscle fascicle lengths.

\subsubsection{Internal structural changes during isometric contractions of the muscle belly}
\begin{figure}[h]
\begin{center}
 \includegraphics[width=\textwidth]{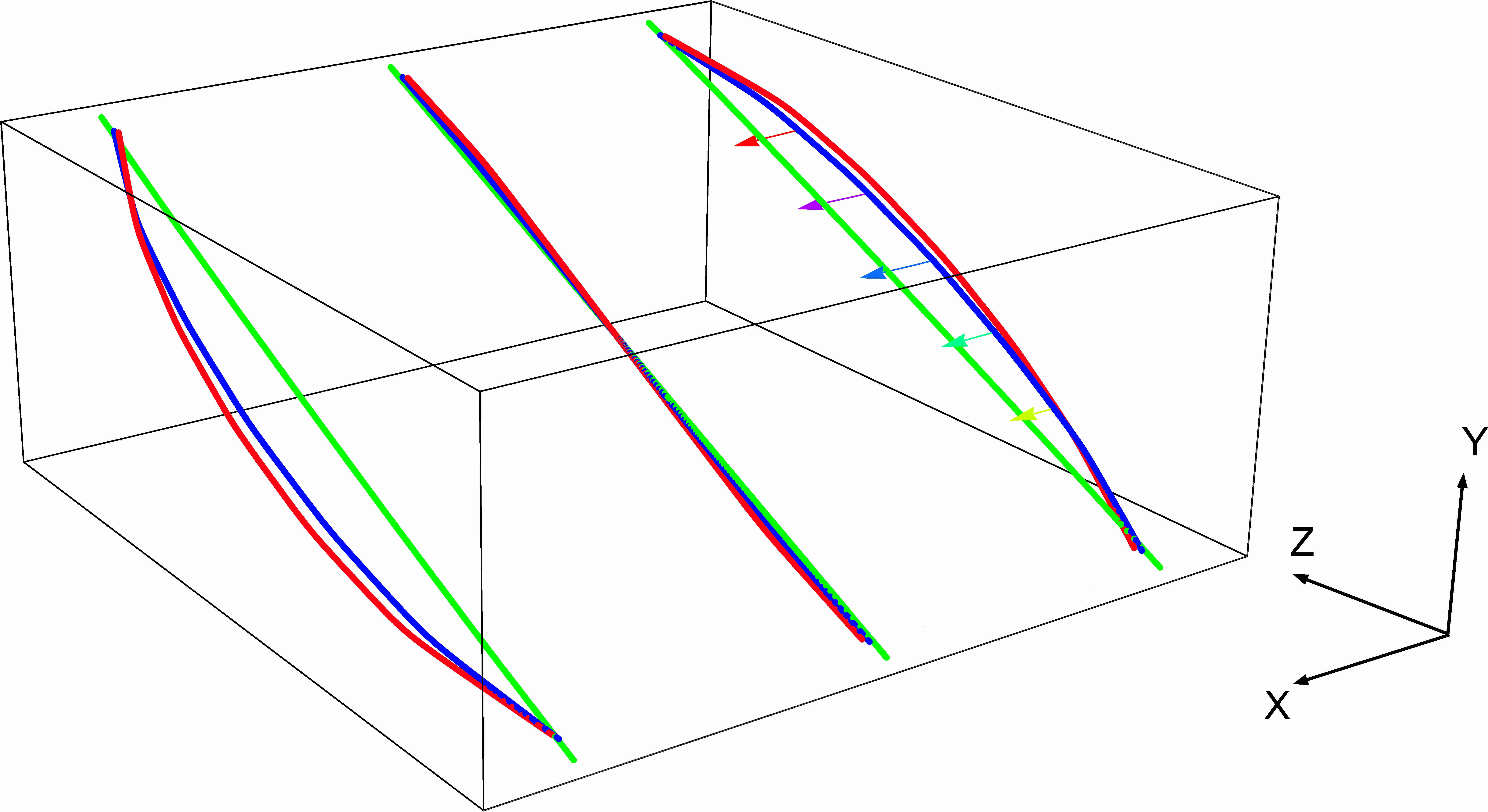} % This is a *.jpg file
\end{center}
\textbf{\refstepcounter{figure}\label{fig:planes} Figure \arabic{figure}.}{ Geometry of the muscle fascicles within the muscle belly (A), shown for their mid-transverse (B) and mid-longitudinal (C) planes. The frames with black fascicle lines are in relaxed state and the frames with red fascicle lines belong to muscle fascicles at a 40\% activity level. The active fascicles show a decrease in thickness and an increase in width in the longitudinal and transverse sections, respectively. Note that the fascicles in the longitudinal section (fascicle plane) are mostly curved to S-shapes in the active state.}
\end{figure}

Both end planes of the aponeuroses for the initial geometry were fixed and the activation was uniformly ramped up. Geometrical properties of fascicles both in  2D (fascicle curvature) mid-longitudinal and transverse planes (Figs. \ref{fig:planes}, \ref{fig:curvemap}) and 3D (fascicle path, along-fascicle and transverse strains) were measured at different activity levels (Fig. \ref{fig:3dpath} and Table \ref{tab:strains}).  Undeformed fascicles (Figs. \ref{fig:planes}, \ref{fig:3dpath}) were chosen as groups of  points that fit along lines that connect the two aponeuroses and have 15 degrees inclination (pennation) in the initial geometry. These fascicles were then tracked throughout all simulations to measure the structural deformations at the fascicle level. The mean pennation and curvature of the fascicles along with the along-fascicle (longitudinal) and transverse strains were extracted from the deformed fascicle data after the contractions had been simulated. The extent of fascicle curvature across the whole muscle belly in its mid-longitudinal plane was quantified by its root-mean-square (RMS) value for each activity level (\% MVC). Fascicle sheets were defined as the 3D faces that run longitudinally through muscle and contain fascicles that were originally in the same YZ-plane of the undeformed geometry. Figure \ref{fig:planes}B shows the intersection of these sheets with the mid-transverse plane.

\subsection{The effect of tendon and aponeurosis properties on structural changes of the muscle tendon unit}
Proximal and distal tendons were attached to the geometry of the muscle belly, where the distal tendon mimics the Achilles tendon. Both tendons had the same thickness and width as aponeuroses, but had lengths of 20 and 160 mm for the proximal and distal tendons, respectively. Initial tests showed that considerable rotations of the muscle belly during contraction as the aponeuroses end planes aligned along the line-of-action of the whole muscle tendon unit (Fig. \ref{fig:unsupported}). To minimize this rotation, the deep aponeurosis (that was attached to the distal tendon) was constrained to not move any more in a deep direction during contraction. The free end of proximal tendon was fixed and the free end of the
distal tendon was pulled about 0.2\% of the total muscle-tendon unit length as an initialization  step to settle the system into a initially stable structure. It was then fixed to keep the muscle-tendon unit isometric. Two situations were investigated: (1) the tendon and aponeurosis had the same material properties that were equal to the tendon properties, and (2) the tendon and aponeurosis had distinctive material properties which are shown below. These simulations achieved a common activation level of 10\%, and the patterns of aponeurosis and tendon strains were compared for the two material formulations.

 Constitutive equations for the tendon and aponeurosis. The mathematical formulation and implementation of these properties can be found in \cite{rahemi2014regionalizing}. We denote by $\lambda$ and $\sigma$ are along-fascicle stretch and stress, respectively, and $I_1$ is the first invariant of right Cauchy-Green deformation tensor.s 

For the tendon, the along-fascicle stress-stretch (in Pa) is given by  
 \begin{equation} \label{equ:tend_stress} \sigma_{Tend}(\lambda)=\begin{cases} 
      10^4 \times 1.904\times(\lambda ^{68.8} -1), & 1 \leq \lambda \leq 1.07 \\
     10^4 \times 1.904\times(6758\times(\lambda-1.07)+104.1), &  1.07< \lambda. 
    \end{cases} \end{equation}
 The tendon base (matrix) material strain energy (Pa) is given by 
            \begin{equation} \label{equ:base_tend}
\Psi_{Tend}= 10^4 \times 2.857\times(I_1-3). 
\end{equation}             
For the Aponeurosis, the along-fascicle stress-stretch properties are given by  \begin{equation} \label{equ:apol_stress} 
\sigma_{Apo\ }(\lambda)=\begin{cases} 
    10^6 \times 3.053\times(\lambda ^{124.6} -1), & 1 \leq \lambda \leq 1.025 \\
    10^6 \times 3.053\times(17375 \times(\lambda-1.025)+20.7), &  1.025< \lambda. 
   \end{cases}
\end{equation}  while the base material of the aponeurosis is given by  \begin{equation*} \label{equ:base_tend}
\Psi_{Apo\ }= 10^4 \times 57.84\times e^{579.6\times(I_1-3)} . 
\end{equation*}

\section{Results}
The force-length properties for the contracting muscle belly are shown in Figure \ref{fig:fl} along with selected data from experimental studies on human muscle. As the muscle was activated, the stretch in the connective tissues allowed the fascicles to shorten, and so the fascicle lengths were different between the active and passive states. Plots shown in Figure \ref{fig:fl} are all for equivalent fascicle lengths, and so the active force was calculated by subtracting the passive force at a slightly longer belly length away from the total force for a contracting muscle. The total and active muscle belly force showed a peak for fascicle stretch of 1, however, the overall shapes of the active and passive plots for the muscle belly were different from the plots for purely muscle fascicles due to the effects from the aponeurosis, muscle structure and pennation.

This modelling framework has previously shown \cite{rahemi2014regionalizing} that the belly force and fascicle pennation becomes larger when the activation state of the muscle belly increases. In the current study the pennation also increased when the belly was passively shortened, and decreased when the belly was passively lengthened. The range of pennation for passive and 30\% active belly were 11.6-19.3 degrees and 13.4-21.2 degrees, respectively, as the belly length was reduced.

\begin{figure}[p]
\begin{center}
 \includegraphics[width=\textwidth]{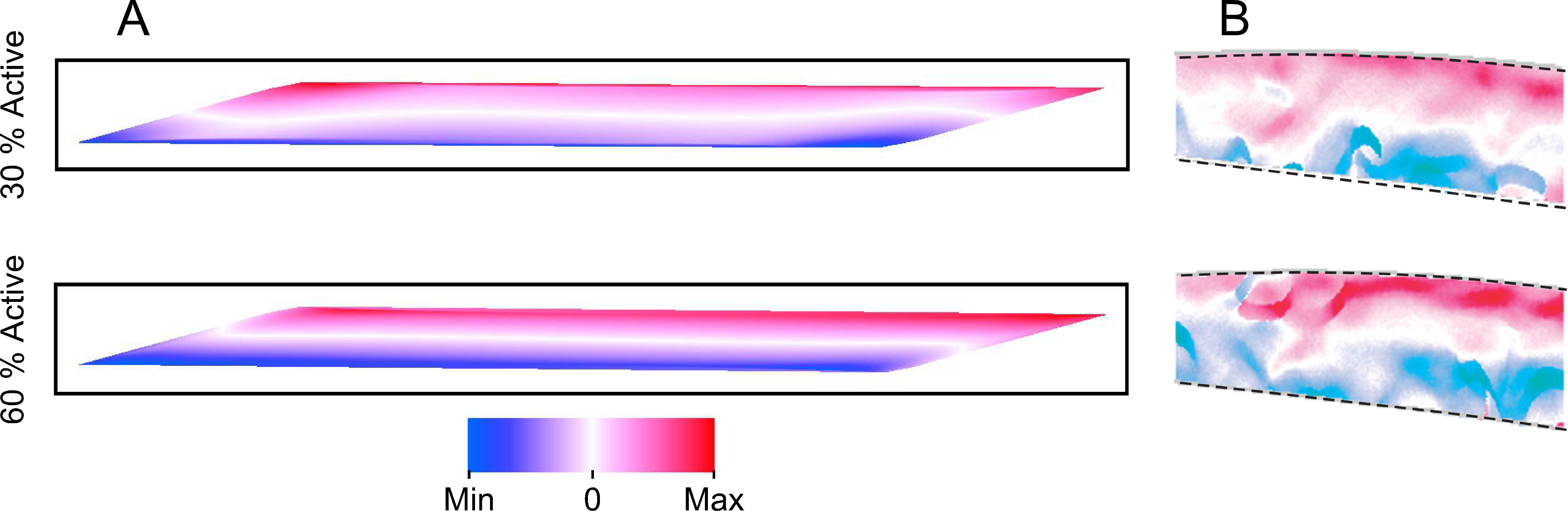} % This is a *.jpg file
\end{center}
\textbf{\refstepcounter{figure}\label{fig:curvemap} Figure \arabic{figure}.}{ Intensity map showing the magnitude of the fascicle curvature for 30 and 60\% activity. Mid-longitudinal plane fascicle curvature map after contraction had been simulated (A). Curvature map for a similar fascicle plane measured in human MG \cite{namburete2011computational} (B).}
\end{figure}

\begin{figure}[p]
\begin{center}
 \includegraphics[width=\textwidth]{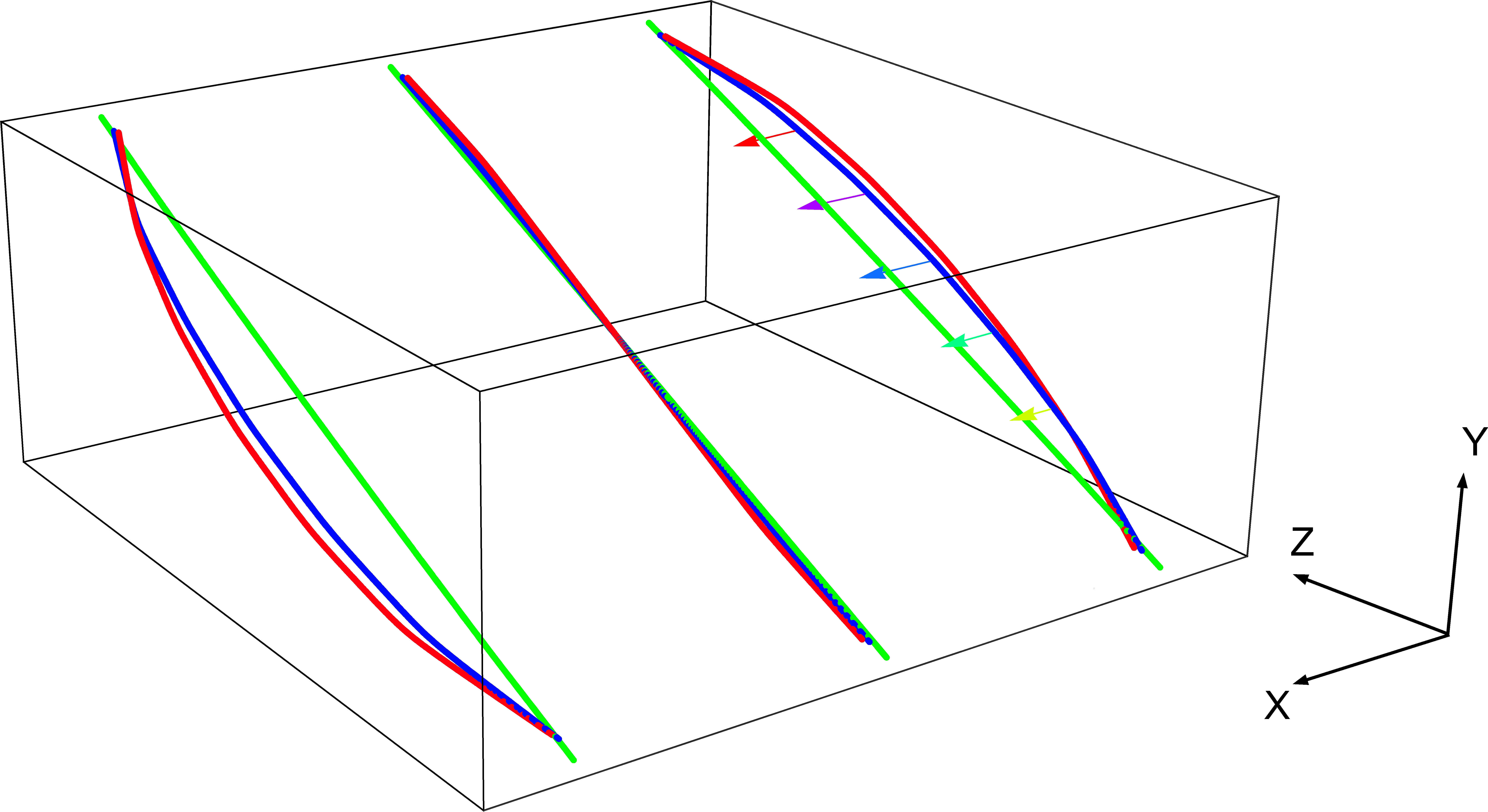} % This is a *.jpg file
\end{center}
\textbf{\refstepcounter{figure}\label{fig:3dpath} Figure \arabic{figure}.}{ 3D paths of three fascicles crossing the mid-transverse plane. Each fascicle is plotted for 0 (green), 30 (blue) and 60\% (red) activity levels. The arrows show the normals to a medial/lateral fascicle at 30\% activity and are coloured by their azimuthal angle where the azimuthal angle is the angle between the projection of the fascicle path in the XY-plane with the X-axis.  The change in azimuthal angle from 80 ( yellow ) to 99 degrees ( red ) shows that the fascicle sheets curve away from the centre of the muscle belly.}
\end{figure}

\begin{figure}[p]
\begin{center}
 \includegraphics[width=\textwidth]{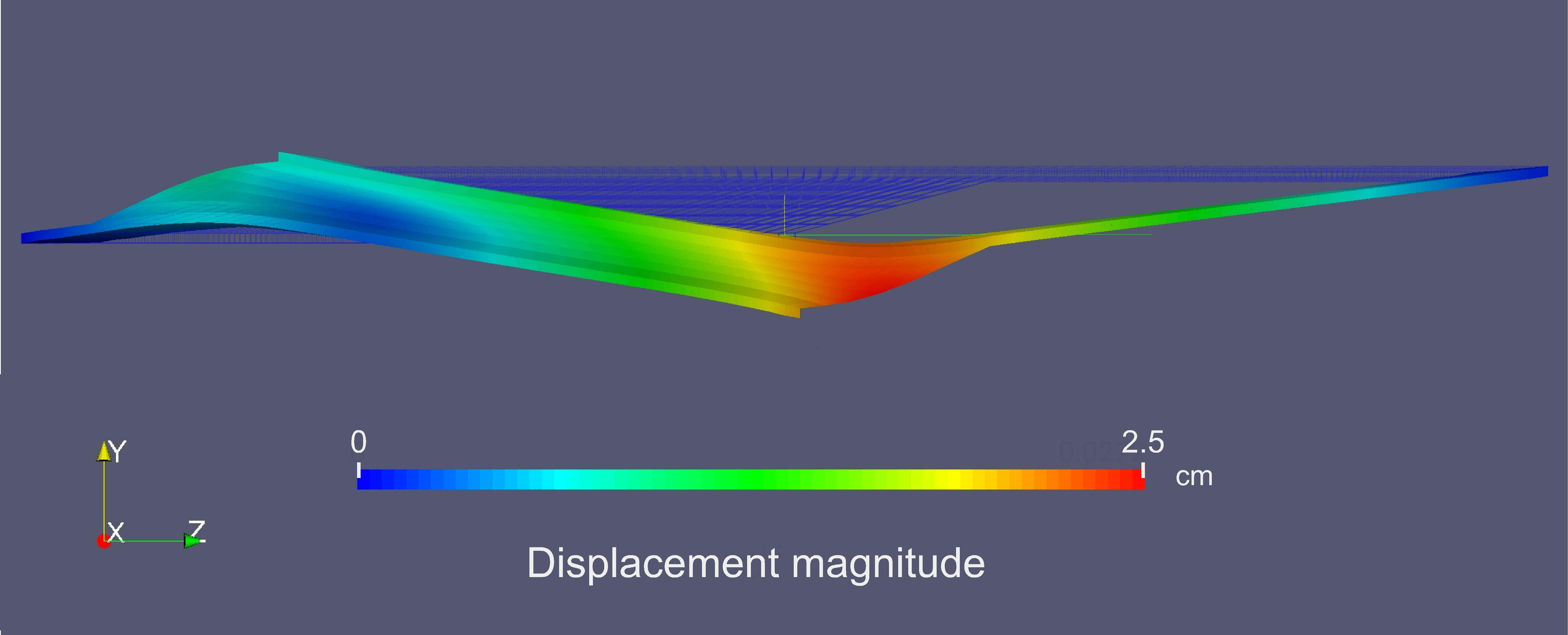} % This is a *.jpg file
\end{center}

\textbf{\refstepcounter{figure}\label{fig:unsupported} Figure \arabic{figure}.}{ Displacement of whole muscle-tendon unit when activated without deep or superficial constraints. }
\end{figure}

\begin{figure}[p]
\begin{center}
 \includegraphics[width=\textwidth]{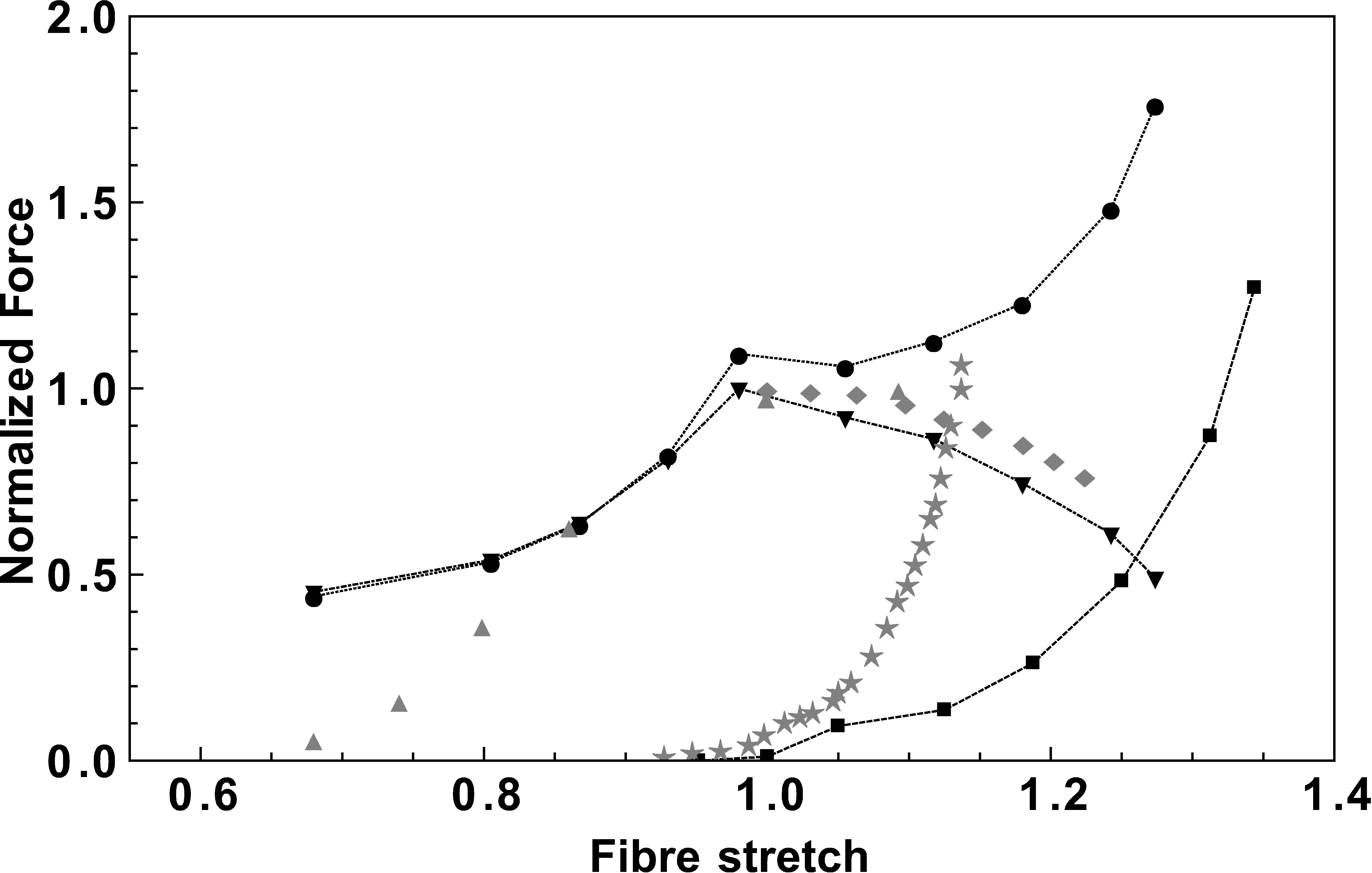} % This is a *.jpg file
\end{center}
\textbf{\refstepcounter{figure}\label{fig:fl} Figure \arabic{figure}.}{ Measured (gray) and modelled (black) force-length properties of human calf muscles. The simulations reached a 30\% activation, and the forces have been normalize to achieve a maximum active force of 1. The black lines without symbols show the active (solid line) and passive (dashed line) force-length properties that were input for the fascicles \cite{rahemi2014regionalizing}. The  black lines  with symbols show normalized active (inverted triangles), passive (squares) and total (circles)  forces for the whole muscle belly. The normalized passive (stars) and active (diamonds) human MG force was measured from twitch contractions \cite{Hoffman2012}. The active human soleus (triangles)  forces were measured from tetanic contractions \cite{maganaris2001vivo}.}
\end{figure}

\begin{figure}[p]
\begin{center}
 \includegraphics[width=\textwidth]{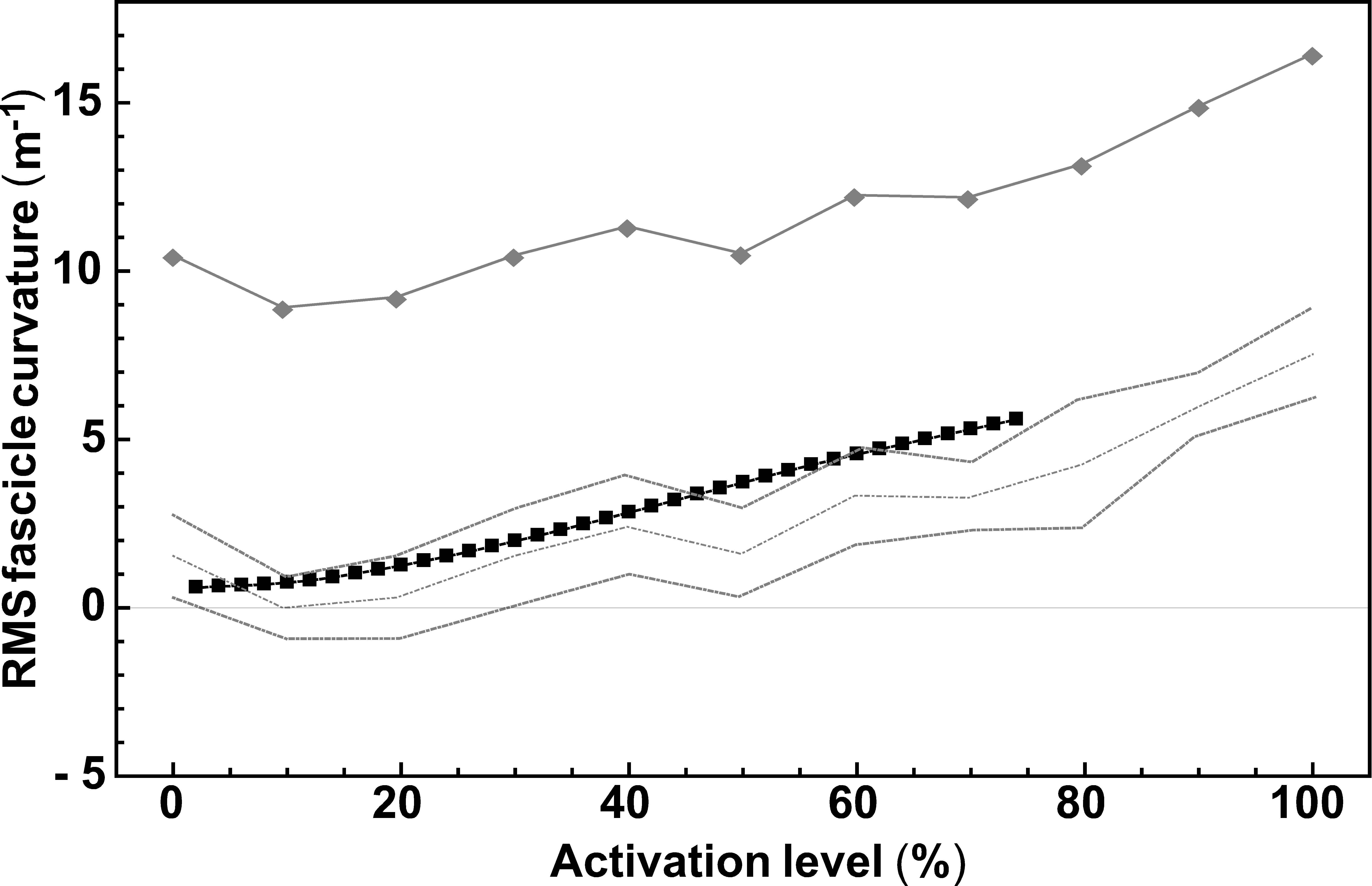} % This is a *.pdf file
\end{center}
\textbf{\refstepcounter{figure}\label{fig:RMS} Figure \arabic{figure}.}{ The root-mean-square curvatures of the fascicles in mid-longituninal plane increased with activation for both simulation (black) and experimental (gray line with symbols)\cite{namburete2011computational} results. The lower gray lines show the change in mean RMS curvature ($\pm$S.D.) from the experimental study.}
\end{figure}

The muscle fascicles in the MG belly, changed from their initially straight configuration to a curved state during contraction. The fascicles showed an S-shaped profile in the mid-longitudinal plane (Fig. \ref{fig:planes}) with the fascicles intersecting with the aponeurosis at a lower angle than their mean orientation would predict. These curvatures profiles match those that we have previously seen experimentally using ultrasound-based imaging \cite{namburete2011computational}, and both are shown in Fig. \ref{fig:curvemap}). The magnitude of the fascicle curvatures increased as the contraction level increased, and the increases in curvature matched the increases experimentally observed in contracting MG (Figs. \ref{fig:curvemap}, \ref{fig:RMS}).

Strain measures for muscle tissue in the centre of the muscle belly are shown for an isometric contraction at 40\% in Table \ref{tab:strains} along with experimentally measured values \cite{wakeling2014transverse}. The transverse strains in the fascicle (mid-longitudinal:YZ) plane were much smaller than the strains normal to this plane. The Poisson's ratio in the fascicle plane was calculated as the magnitude of the ratio between transverse and along-fascicle strains in this plane and was 0.089. 

The fascicle sheets bulged in both medial and lateral directions when the muscle belly contracted (Figs.  \ref{fig:planes}, \ref{fig:3dpath}), and the bulge increased as the activity level rose. The path of the fascicles in 3D showed them running along the fascicle sheets as they bulged, and thus formed a part of a helix (demonstrated by their varying azimuthal angle along their length  (Fig. \ref{fig:3dpath}).   

\begin{figure}[p]
{
 \subfloat[Two tissues simulation]{\label{fig:twot}\includegraphics[width=\textwidth]{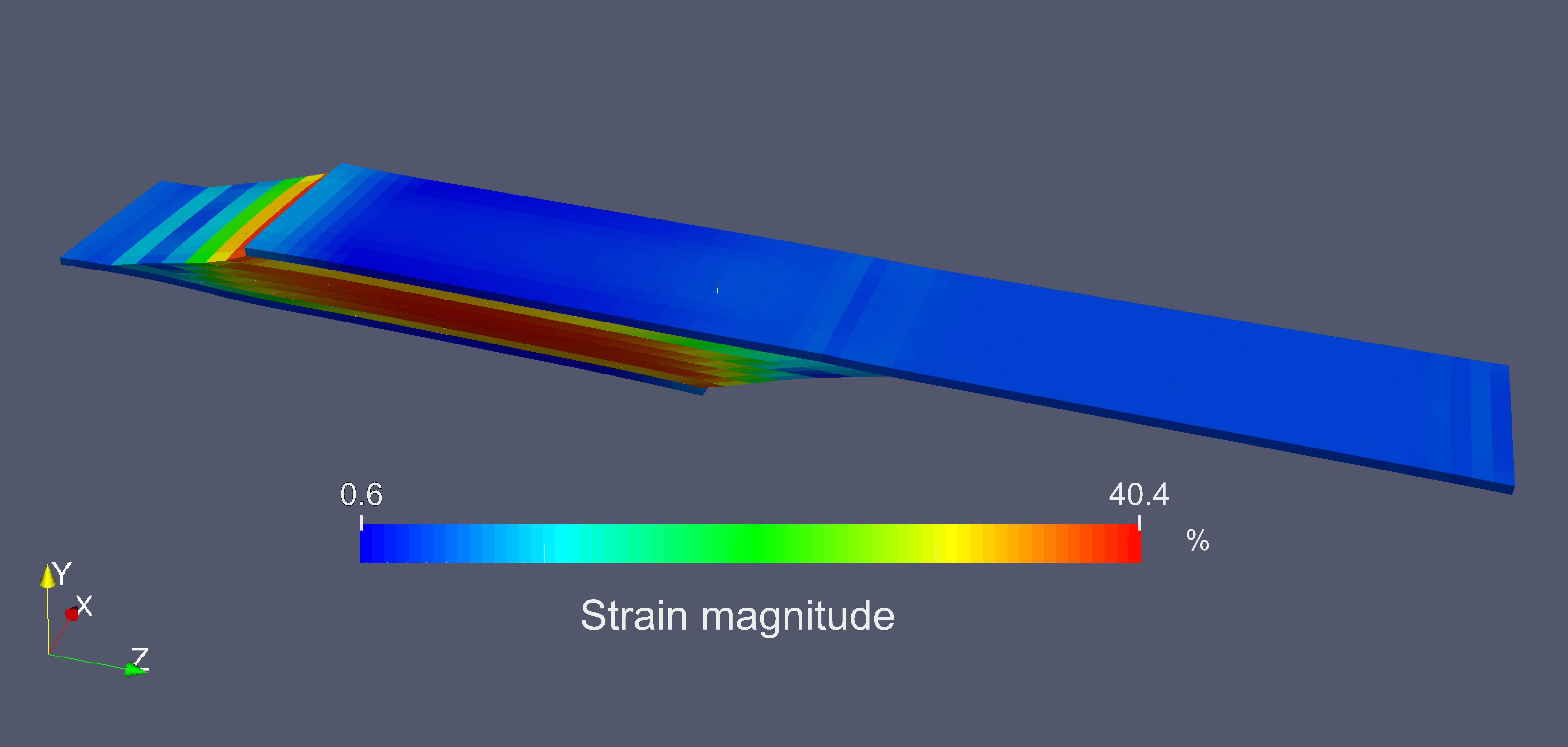}}
 \\
  \subfloat[Three tissues simulation]{\label{fig:threet}\includegraphics[width=\textwidth]{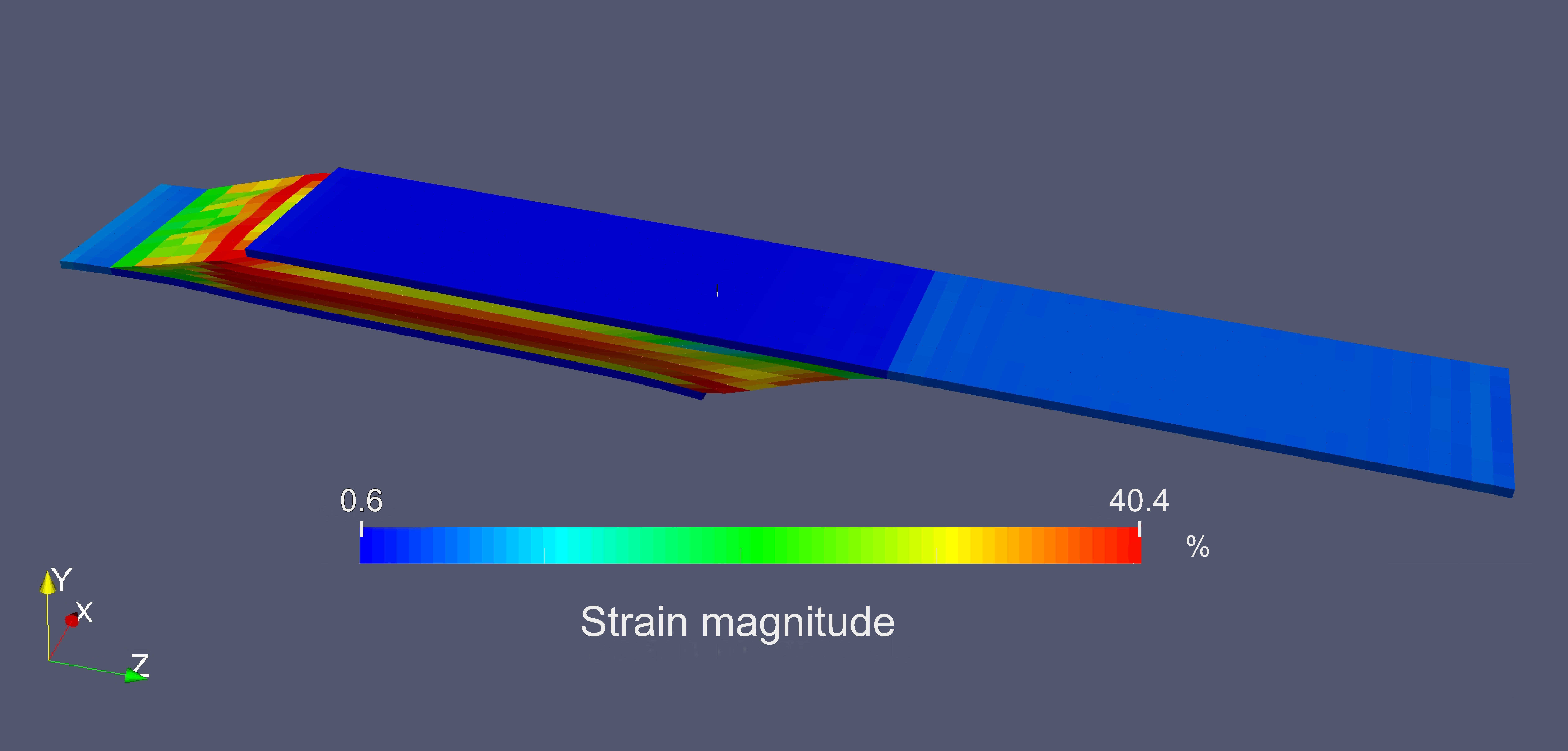}}
  \quad
  \\ }
\textbf{\refstepcounter{figure}\label{fig:multi} Figure \arabic{figure}.}{ Total strain in the muscle tendon unit tissue at a 10\% activity level for two material conditions: equal material properties for aponeurosis and tendon (A), tissue specific properties (Table \ref{tab:materials}) for aponeurosis and tendon (B). }
\end{figure}

When the whole muscle-tendon unit was simulated (with the external tendons included), the muscle belly showed substantial rotations as the aponeurosis end planes aligned to be closer to the line-of-action of the muscle (Fig. \ref{fig:unsupported}). Subsequent simulations of the MTU constrained the deep aponeurosis to not displace any deeper, and this forced the bulging of the muscle belly to be in the superficial direction. This was to emulate a simplified set of constraints that occur on the MG within the intact leg. The final simulations (Fig. \ref{fig:multi}) showed that when a stiffer aponeurosis was used instead of adopting tendon properties, the strains in aponeurosis were smaller. Also the strains in the muscle tissue were more uniform when a stiffer material for the aponeurosis was used.
\section{Discussion}

Validating a mathematical framework and numerical implementation of it for human muscle is a challenge, due in part to the fact that muscle forces cannot be directly measured in vivo. In this study we have compared the force output from a computational 3D FEM model with the forces estimated from studies of ankle joint flexion-extension experiments. The general pattern of the force-length relationship generated by the model matches those from the experimental studies. Experimental measures can identify the overall shape of the muscle with MRI \cite{gilles2006anatomical} and even the internal trajectories of the muscle fascicles using diffusion-tensor MRI \cite{heemskerk2009quantitative,heemskerk2011vivo,infantolino2012arrangement}. While this information is very important, the relatively long scan times of MR imaging preclude such measurements for active contractions \cite{rana2011vivo}. However, the aim of the presented muscle model is to understand the mechanisms occurring during muscle contractions. It is therefore important to validate the muscle model in its contracted state. For this study we have used ultrasound-based measures \cite{namburete2011computational, Rana20133d} of the internal structure during contraction (fascicle orientations, curvatures, and strains) to validate the model.

The model in this study has a simplistic initial geometry that has the overall dimensions and mean fascicle pennation of the MG in man, but without the details of the geometry or internal structure. Furthermore, all the muscle fascicles within the model had the same material properties and thus represented the same fibre-types. Additionally, the activation was uniform across all fascicles: again these are gross simplifications compared to the physiological complexities and variations that occur within muscles \textit{in-vivo}. Nonetheless, the emergent features from the model showed a remarkable similarity to the experimental measures that are available for comparison, giving confidence that the model can identify general features and consequences of the muscle structure that were not a result of idiosyncrasies or muscle-specific details of geometry, structure or activation.

Intramuscular pressure develops within muscles during contraction \cite{Sjogaard1986,sejersted1984intramuscular,Maton2006}, and the fascicles curve around the regions of higher pressure. Previous modelling studies \cite{Van_Leeuwen1992,vanLeeuwen1995} have shown how the curvatures in both the muscle fascicles and aponeurosis must balance the intramuscular pressure, and indeed our current model shows curvatures developing in both these structures. However, in these previous studies the curvatures of the muscle fascicles were constrained to be constant along their lengths, whereas this was not a constraint in the current model. The muscle fascicles in the current model started straight in their initial configuration, but developed S-shaped profiles when quantified in the mid-longitudinal plane. Both the S-shaped profiles and the increases in curvature that occurred with increasing activity and muscle force mirror those that we have previously imaged for the MG using B-mode ultrasound \cite{namburete2011computational,Rana20133d}. A consequence of the S-shaped trajectories is that the angle at which the fascicles insert onto the aponeurosis can be reduced, allowing for a greater component of traction in the line of action of the whole muscle along the direction of the aponeuroses. 

When tracked in 3D, the muscle fascicles followed curved paths on their fascicle sheets indicating that change in architecture is not simply due to a bulge of the sheets. The active configuration  of these fascicles indicate that S-shaped fascicles in 2D curvature maps (Fig. \ref{fig:curvemap}) are not only the result of projecting the fascicles on a 2D plane {rana2014curve} but comes from curling of the fascicles in a helical path. These 3D helical paths are curved around the centre of the  muscle (Fig. \ref{fig:3dpath}) where the intramuscular pressure is higher.

It is generally assumed that muscle fascicles are isovolumetric \cite{Baskin1967}, and isovolumetric assumptions dictate the relation between longitudinal and transverse strains. Poisson's ratio is the absolute value of ratio of the transverse to the longitudinal strain, and should be 0.5 for small strains in an incompressible and elastic material. The simulations in this study showed that as the activation increased, the transverse strain (in the mid-longitudinal plane) was lower than expected, resulting in a Poisson’s ratio of 0.089, however this was compensated for by greater transverse strains in the orthogonal direction (Table \ref{tab:strains}). The muscle fascicles were represented as transversely isotropic materials in this model \cite{rahemi2014regionalizing}, and so the asymmetry in their transverse bulging must reflect asymmetries in the transverse stresses acting on the fascicles. Being a unipennate model, there would have been a larger compressive force in the mid-longitudinal plane that was bounded by the aponeuroses that were being squeezed together by the pennate fascicles, than in the medial-lateral direction where there was no aponeurosis bounding the muscle. Indeed, the model has showed muscle belly bulging to its sides, but decreasing in its thickness between the aponeuroses during contraction \cite{rahemi2014regionalizing}, in a similar manner to the decreases in thickness observed for the MG in vivo \cite{randhawa2013-1}. Recently we have quantified transverse bulging of the muscle fascicles in the MG from B-mode ultrasound images \cite{wakeling2014transverse}, showing a Poisson’s ratio of 0.09; this matches the simulated results and provides confidence that emergent features of the model explain realistic features of muscle contraction. 

When the model was evaluated with external tendons, there was a need to constrain displacements of the  geometry since the unconstrained simulation (Fig. \ref{fig:unsupported}) showed a large displacement of the muscle in the Y-direction. This illustrates that a range of additional boundary constraints may need to be applied to finite element models of muscle-tendon units in order to result in more realistic deformation. 

In the case that the aponeurosis and tendon were given the same material properties a pattern of non-uniform  strains resulted in the aponeurosis. This non-uniformity in strain is similar  to that observed in previous experiments \cite{finni2003nonuniform,Muramatsu2001}, but our modelling study shows this can be an emergent feature of the muscle, and not necessarily due to differnces between active and inactive motor units in submaximally activated muscle, as previously suggested \cite{finni2003nonuniform}. The aponeurosis strains were smaller than the tendon strains for both formulations (Table \ref{tab:materials}) of material property (Fig.\ref{fig:multi}). Although there is an obvious jump in strain between the tendon and aponeurosis when a stiffer material is used for the aponeurosis, the difference in strains was less than 2\%. A benefit of such a material distribution would be that a more uniform distribution of strains occurs in the fascicles, and this would allow the fascicles to produce more consistent forces along their length.

The simulated results from this finite element model match the general patterns from experimental and imaging results. Whole muscle force is partly shaped by the internal geometry of the muscle fascicles, and their interactions with the aponeuroses \cite{rahemi2014regionalizing}, and so cannot be explained entirely by modelling a muscle as a scaled-up muscle fibre \cite{wakeling2011modelling}. As the fascicles shorten they must increase in cross-sectional area in order to maintain their volume, but asymmetric bulging occurs due to asymmetries in the compressive stress acting on the fascicles during contraction. The fascicles curve and adopt S-shaped profiles that align their traction to be closer to the aponeurosis direction, and they curl across fascicle sheets that in turn bulge around the intramuscular pressure that develops during contraction. Material properties of the aponeuroses affect the strains in the fascicles and thus their force generating potenitial. The muscle model that we have validated in this study will provide a useful tool for understanding the mechanisms that relate muscle structure to its contractile function.   

% \usepackage{booktabs}
% \usepackage{graphicx}
% \usepackage{booktabs}

% \usepackage{booktabs}
\section*{Acknowledgement}
We gratefully acknowledge funding from Natural Sciences and Engineering Research Council of Canada (Nilima Nigam and James M. Wakeling) and the Canada Research Chairs Program (Nilima Nigam).

\bibliographystyle{plain}
\bibliography{refjbio.bib}

\section*{Figures}
Figures and captions.
\\
\\
\\
\\
\\
\\
\\

\section*{Tables}
\label{sec:tabs}
Tables and captions.
\\
\\
\\
\\
\\
\\
\\

\begin{table}[p]

\textbf{\refstepcounter{table}\label{tab:strains} Table \arabic{table}.}{ Along-fascicle and transverse strains for fascicles in the middle of the muscle belly for 40\% activity (Fig.\ref{fig:planes}). The Poisson's ratio in the mid-longitudinal plane is calculated as the magnitude of the ratio of the transverse (cross-fascicle) to the along-fascicle strain. The last row shows the measured Poisson's ratio from 2D ultrasound images in the mid-longitudinal plane of the MG during dynamic contractions \cite{wakeling2014transverse}.\\ \\}
\resizebox{\textwidth}{!}{%

{\tiny
\begin{tabular}{ll}

\toprule
\textbf{Parameter}                                     & \textbf{Values} \\ \midrule
Along-fascicle (longitudinal) strain (\%) - simulation                     & -8.22           \\
Transverse (cross-fascicle) strain (\%) - simulation      & 0.74            \\
Out of plane (width) strain (\%)- simulation & 7.83            \\
Mid-longitudinal plane Poisson's ratio - simulation              & 0.089           \\
Poisson's ratio - \textit{in-vivo} \cite{wakeling2014transverse}      & 0.09$\pm$0.01   \\ \bottomrule
\end{tabular}}}
\end{table}

\end{document}